\documentclass[letterpaper, 10 pt, conference]{ieeeconf}  

\IEEEoverridecommandlockouts                              

\let\labelindent\relax
\usepackage[hidelinks]{hyperref}
\usepackage[noadjust]{cite}
\usepackage{amsmath,amssymb,amsfonts, lipsum, mathtools, cuted, enumitem}
\usepackage{graphicx}
\usepackage{optidef} 

\usepackage[open,openlevel=1]{bookmark}

\usepackage{ntheorem}
\usepackage{thmtools}
\newtheorem{thm}{Theorem}

\newtheorem{prop}{Proposition}
\newtheorem{assum}{Assumption}
\newtheorem{rem}{Remark}
\newtheorem{prob}{Problem}

\allowdisplaybreaks

\title{\LARGE \bf
Uncertainty Learning for LTI Systems with Stability Guarantees 
}

\author{Farhad Ghanipoor$^{1}$, Carlos Murguia$^{1}$, Peyman Mohajerin Esfahani$^{2}$, and Nathan van de Wouw$^{1}$
	\thanks{$^{1}$ Farhad Ghanipoor, Carlos Murguia,  and Nathan van de Wouw are with the Mechanical Engineering Department, Eindhoven University of Technology, Eindhoven, The Netherlands.  {Emails: \tt\small f.ghanipoor@tue.nl}, {\tt\small C.G.Murguia@tue.nl}, and {\tt\small N.v.d.Wouw@tue.nl}.}%
	\thanks{$^{2}$ Peyman Mohajerin Esfahani is with the Delft Center for Systems and Control, Delft University of Technology, Delft, The Netherlands. {Email: \tt\small P.MohajerinEsfahani@tudelft.nl}.}%
}

\begin{document}

	\maketitle
	\thispagestyle{empty}
	\pagestyle{empty}

	\begin{abstract}
		  We present a framework for learning of modeling uncertainties in Linear Time Invariant (LTI) systems. We propose a methodology to extend the dynamics of an LTI (without uncertainty) with an uncertainty model, based on measured data, to improve the predictive capacity of the model in the input-output sense. The proposed framework  guarantees stability of the extended model. To achieve this, two semi-definite programs are provided that allow obtaining optimal uncertainty model parameters, given state and uncertainty data. To obtain this data from available input-output trajectory data, we introduce a filter in which an internal model of uncertainty is proposed. This filter is also designed via a semi-definite program with guaranteed robustness with respect to uncertainty model mismatches, disturbances, and noise. Numerical simulations are presented to illustrate the effectiveness and practicality of the proposed methodology in improving model accuracy, while warranting model stability.
	\end{abstract}

	\section{INTRODUCTION} \label{sec: introduction}
	
	Dynamical systems modeling has been a key problem in many engineering and scientific fields, such as biology, physics, chemistry, and transportation. For modeling of dynamical systems, physics-based approaches can be used to construct generic mathematical models by using first-principles fundamental laws. When modeling dynamical systems, it is of key importance to use such well-established principles of physics and other prior system knowledge such as stability \cite{revay2023recurrent}. However, many complex and uncertain systems in real life possess partially known physics only \cite{abbasi2022physics}. 
	 Even for systems in which accurate physics-based models are known, such as high-tech systems and robotics, there are unavoidable (parametric and non-parametric) uncertainties that affect the model's predictive capacity.
	
	For a class of linear uncertain dynamical systems, this paper focuses on learning models for uncertainties while guaranteeing stability of extended models (prior models plus uncertainty characterization), given available input-output data. This problem contrasts with black box modelling approaches, e.g., using Neural Networks (NNs) or Gaussian Processes (GPs), as we incorporate prior relations that come from first-principles into the modeling and learning scheme.
	
	\emph{Existing Literature:} Our approach is fundamentally different from well-known existing Physics-Informed (PI) learning techniques where standard black box models (such as NNs or GPs) are trained constrained to satisfy physics-based relations \cite{raissi2019physics}. Yazdani et al. in \cite{yazdani2020systems} use this technique to construct the so-called Physics Informed Neural Networks (PINNs). They constrain a known physics-based model during the training of a NN-based model (i.e., penalize loss function for the mismatch between the physics-based model and the NN as a soft constraint) and incorporate physics knowledge in the structure of the NN. 
	 Although this method provides a NN as a system model (as well as parameters for physics-based models), it does not give a closed-form expression for uncertainty in the physics-based model (i.e., it does not account for modeling mismatches due to unmodeled physics). Furthermore, parameters of both physics-based models and NNs are learned simultaneously, which increases the computational burden.

	The approach proposed in this paper also differs from the so-called Sparse Identification of Nonlinear Dynamics (SINDy) scheme \cite{brunton2016discovering}. SINDy assumes full knowledge of system states and their time derivatives. Then, based on known physics-based models and known variables (i.e., system states and their derivatives) a library of functions is generated that can be incorporated in dynamical models to account for uncertainty. To select the active functions in models, sparse identification algorithms are exploited. This approach has demonstrated quite accurate performance in sparse model identification of complex nonlinear systems \cite{champion2019data, champion2020unified, loiseau2018constrained}. However, SINDy not only requires full-state measurement but also requires the derivative of states to be known. Although the state derivatives can be approximated numerically if the complete state is known, most numerical methods are noise sensitive. Furthermore, the requirement of full-state measurements is a strong assumption for most dynamical systems. In our work, we do not require measurements of the full-state and its time derivative. The proposed algorithms need input-output data only.  
	
	Our approach, augments a known physics-based model by a black-box model used as a correction term. Such generic approach is also taken by Quaghebeur et al. in \cite{quaghebeur2021incorporating}, who add an NN model to a known physics-based model with unknown parameters. This approach allows maintaining the basic structure of the model that comes from first principles, which improves interpretability. However, for training of the NN and the physics-based model parameters, they need to simulate the hybrid model during the training process, which is computationally expensive. Furthermore, the main drawback of this method is that it assumes that the initial state of the dynamic system is known or at least it requires measuring all the states (full-state measurement) of the true dynamical system. 
	
	Furthermore, our approach offers stability guarantees for the extended LTI model (i.e., the model consisting of the known physics-based model and the uncertainty model). The identification of stable LTI models has (mainly) been studied in the context of discrete systems \cite{umenberger2018maximum, tobenkin2010convex, abbasi2022data}. For instance, in \cite{manchester2021contraction}, the authors provide convex constraints to ensure incremental stability for linear non-autonomous discrete-time models and some nonlinear models such as recurrent neural networks. Additionally, there exist studies that have explored model identification with asymptotic stability guarantees for discrete LTI systems using subspace identification methods \cite{van2001identification, miller2013subspace}. Subspace identification methods involve obtaining an estimate of state sequence or extended observability matrix, followed by solving a least squares problem to estimate the model parameters. One way to ensure the asymptotic stability of the model is to add stability constraints to the least squares optimization \cite{lacy2003subspace}. This addition results in a convex linear program with mixed equality, quadratic, and semi-definite constraints. Moreover, there exist some studies such as \cite{pillonetto2010new, scandella2022kernel} which provide non-parametric model identification with stability guarantees for discrete and continuous LTI systems using kernel-based approaches. It can be observed that stable LTI model identification has a rich literature, with a focus on identifying the complete dynamics. In this context, our focus is on identifying the missing elements (uncertainty) in known physics-based models.
	
	In this paper, we propose a framework for learning of modeling uncertainties in physics-based models applicable to Linear Time Invariant systems (LTIs). We first focus on fitting uncertainty models, assuming that some realizations of input, (estimated) uncertainty, and (estimated) state are given, while guaranteeing asymptotic stability of the extended model (i.e., known physics-based model plus uncertainty model). This is achieved by formulating the problem as a constraint supervised learning problem. 
	
    One key challenge in this problem is the introduction of stability constraints, which is addressed using Lyapunov-based tools. The stability criteria typically lead to a non-convex optimization problem. We tackle this challenge by proposing two different approaches:
	\begin{enumerate}
	\item \textbf{Cost Modification:} The first approach involves a change of variables, resulting in cost function being rewritten in terms of these new variables (Theorem \ref{theorem:learning_modified_cost}). 

	\item  \textbf{Constraint Modification:} The second approach introduces a sufficient condition to fulfill the stability constraint by solving a convex program (Theorem \ref{theorem:learning_modified_constraint}).  
	\end{enumerate}
	Having addressed the non-convexity challenge, the paper proceeds to discuss the practical implementation of the framework. Specifically, it outlines a method for estimating uncertainty and state trajectories using input-output data and the known physics-based model (Proposition \ref{theorem:optimal_estimator}). In this context, we treat uncertainty as an unknown term affecting the system dynamics and estimate uncertainty and state using robust state and unknown input observers \cite{ghanipoor2022ultra, ghanipoor2023robust}. \newline
	
	\textbf{Notation:} 
	The symbol $\mathbb{R}^{+}$ denotes the set of nonnegative real numbers. The $n \times n$ identity matrix is denoted by $I_n$	or simply $I$ if $n$ is clear from the context. Similarly, $n \times m$ matrices composed of only zeros are denoted by ${0}_{n \times m}$ or simply ${0}$ when their dimensions are clear. First and second time-derivatives of a vector $x$ are expressed as $\dot{x}$ and $\ddot{x}$, respectively. For $r^{th}$-order time-derivatives of a vector $x$, the notation $x^{(r)}$ is adopted. A positive definite matrix is denoted by $X \succ 0$ and positive semi-definite matrices are denoted by $X \succeq 0$. Similarly, for a negative definite $X \prec 0$ is used, and $X \preceq 0$ for negative semi-definite matrices. The imaginary unit $j$ is defined by $j^2 = -1$. For a transfer function $T(s)$, with $s \in   \mathbb{C} $, $\sigma_{\max}\big(T(s) \big)$ denotes the maximum singular value, and $T^H (s)$ represents the Hermitian transpose. The notation $\text{col}[x_1, \ldots , x_n]$ stands for the column vector composed of the elements $x_1,\ldots,x_n$. This notation is also used when the components $x_i$ are vectors. For a differentiable function $V: \mathbb{R}^{n}  \to \mathbb{R}$ we denote by $\frac{\partial V}{\partial x}$ the row-vector of partial derivatives and by $\dot{V}(x)$ the total derivative of $V(x)$ with respect to time (i.e., $\frac{\partial V}{\partial x} \frac{dx}{dt}$). The notation $tr(W)$ stands for trace of a matrix $W$.  We often omit time dependencies for notation simplicity.

	\section{Problem Formulation} \label{sec: problem formulation}
	Consider the system
	\begin{equation} \label{eq:sys}
		\left\{\begin{aligned}
			\dot{x}_s  =& A x_s+ B_u u+ S_\eta \eta (x_s, u) + B_\omega \omega,\\
			y_s=& C x_s + D_\nu \nu,
		\end{aligned}\right.
	\end{equation}
	where $t \in \mathbb{R}^{+}$, $x_s \in {\mathbb{R}^{n}}$, $y_s \in {\mathbb{R}^{{m}}}$, and $u \in {\mathbb{R}^{{l}}}$ are time, system state, measured output and known input vectors, respectively, and function $\eta: \mathbb{R}^{n} \times \mathbb{R}^{l} \to \mathbb{R}^{n_\eta}$ is unknown modeling uncertainty. Signals $\omega: \mathbb{R}^{+}  \to \mathbb{R}^{n_\omega}$ and $\nu: \mathbb{R}^{+}  \to \mathbb{R}^{m_\nu}$ are unknown bounded disturbances; the former with unknown frequency range and the latter with high-frequency content (e.g., related to measurement noise). Known matrices $(A, B_u, S_\eta, B_\omega, C, D_\nu)$ are of appropriate dimensions, with ${n},{m}, l, n_{\eta}, n_{\omega}, m_\nu \in \mathbb{N}$. Matrix $S_\eta$ is used to indicate in which equation(s) the uncertainty $\eta$ appears explicitly.
	  
	We aim to fit a data-based model for the uncertainty (i.e., $\eta(\cdot)$ in \eqref{eq:sys}) using a supervised learning method, while guaranteeing model stability, with the goal of constructing a more accurate system model (valid at least for trajectories close to the training data set). The proposed methods in Section \ref{sec: learning_sol} assume that a data-set (labeled data) of input, (estimated) uncertainty, and (estimated) state realizations are given. This assumption can be considered as another problem for which a solution is provided in Section \ref{sec: uncertainty_state_estimation}. In what follows, we formulate the problem of uncertainty model learning with stability guarantees. 
	
	For the system in \eqref{eq:sys}, consider the following LTI model 
	\begin{equation} \label{eq:model}
		\begin{aligned}
			\dot{x} &= A x+ B_u u + S_{\eta_l} \eta_l(x, u), \\ 
			  \quad \eta_l(x,u) &:= \Theta_l x + B_l u,
		\end{aligned}
	\end{equation}
where $x \in {\mathbb{R}^{n}}$ is model state and function $\eta_l: \mathbb{R}^{n} \times \mathbb{R}^{l} \to \mathbb{R}^{n_{\eta_l}}$ is the uncertainty model that is parameterized by $\Theta_l, B_l$. Matrices $(\Theta_l, B_l, S_{\eta_l})$ are of appropriate dimensions, with $n_{\eta_l} \in \mathbb{N}$. Matrix $S_{\eta_l}$, similar to $S_\eta$ in \eqref{eq:sys} shows explicit appearance of the uncertainty model $\eta_l$ in the right-hand side and could be different from $S_{\eta}$.

	Next, we define a cost function for supervised learning and the stability constraint. 
	
		\subsection{Cost Function} 
	Let us define the following (given) $i$-th sample (in time) data vector
	\begin{equation*}
		\begin{aligned}
			d_i := {\left[\begin{array}{ccc}
					\hat{x}_i^\top & u_i^\top  & \hat{\eta}_i^\top 
				\end{array}\right]^\top},
		\end{aligned}
	\end{equation*}
	where $\hat{x}_i, u_i,$ and $\hat{\eta}_i$ correspond to given $i$-th realizations of state estimation, input, and uncertainty estimation, respectively. Given $N$ samples of data realizations, define the data matrix $D$ as follows:
	\begin{equation} \label{eq:data_matrix}
		D := \sum_{i = 1}^{N} d_i d_i^\top.
	\end{equation}
Further, define the error vector between the uncertainty model and its (given) estimation as  
	\begin{equation*} \label{eq:error}
		\begin{aligned}
			e_i := \eta_l(\hat{x}_i, u_i)  - \hat{\eta}_i = T d_i
		\end{aligned}
	\end{equation*}
	with
	\begin{equation} \label{eq:T}
		T := {\left[\begin{array}{ccc }
				\Theta_l  & B_l & -I
			\end{array}\right]}.
	\end{equation} 
	Then, we define the following quadratic cost function to be minimized to identify $\Theta_l$ and $B_l$:
		\begin{equation} \label{eq:quadratic_cost}
			\begin{aligned}
				J := \sum_{i = 1}^{N} e_i^\top e_i = \sum_{i = 1}^{N} d_i^\top T^\top T d_i.  
			\end{aligned}
		\end{equation}
	\subsection{Stability Constraint} 
	We aim to formulate a constraint to satisfy asymptotic stability of the model in \eqref{eq:model} via Lyapunov-based stability analysis. 
	
	Consider the quadratic function $V(x) = x^\top P x$ for a positive definite matrix $P \succ 0$. If we can find a $P$ such that $\dot{V} < 0$ for $u=0$ along trajectories of \eqref{eq:model}; then, the model in \eqref{eq:model} is asymptotically stable. The condition $\dot{V} < 0$, for $u=0$ can be stated as 
	\begin{equation} \label{eq:vdot_cond_P}
			(A+S_{\eta_l} \Theta_l)^{\top} P+P (A+S_{\eta_l} \Theta_l) \prec 0,
	\end{equation}
	or equivalently, by applying the congruence transformation of $Q := P^{-1}$, \eqref{eq:vdot_cond_P} can be written as  
	\begin{equation} \label{eq:vdot_cond_Q}
			(A+S_{\eta_l} \Theta_l) Q+Q (A+S_{\eta_l} \Theta_l)^{\top}  \prec 0.
	\end{equation} 
	
	Note that the asymptotic stability conditions above requires the linear matrix of the dynamics $A+S_{\eta_l} \Theta_l$ to be Hurwitz. For a nonzero $u$, this condition implies Input-to-State Stability (ISS) of the model in \eqref{eq:model}  \cite[Col. 5.2]{khalil2002nonlinear}, for any $B$ and $B_l$. Now, we can state the problem we seek to solve. 
		
	\begin{prob}\emph{\textbf{(Uncertainty Model Learning with Stability Guarantee)}} Consider a given data-set of input and (estimated) uncertainty and state realizations. Find the optimal parameters $\Theta_l$ and $B_l$ of uncertainty model $\eta_l(\cdot)$ of the form in \eqref{eq:model} that minimizes the cost function $J$ in \eqref{eq:quadratic_cost}, such that the system model in \eqref{eq:model} is asymptotically stable (i.e., respecting the constraint in \eqref{eq:vdot_cond_P} or \eqref{eq:vdot_cond_Q}). In other words, find the optimal parameters of the following optimization problem:
\begin{mini}|s|[4]           	
	{P,\Theta_l, B_l}                            
	{J}   								
	{\label{eq:nl_sdp_qp}}             								
	{}                           								
	\addConstraint{(A+S_{\eta_l} \Theta_l)^{\top} P+P (A+S_{\eta_l} \Theta_l)}{\prec 0}   					  	
	\addConstraint{P}{\succ 0.}   					  	
\end{mini}
		\label{prob:uncertainty_model_train}
	\end{prob}

	In what follows, we provide an approximate solution to Problem \ref{prob:uncertainty_model_train}.
	
	\section{Approximate Solution to Problem \ref{prob:uncertainty_model_train}} \label{sec: learning_sol}
     The challenge is that the stability condition \eqref{eq:vdot_cond_P}  that appears in the optimization problem \eqref{eq:nl_sdp_qp} is not convex in $P$ and $\Theta_l$. Therefore, in what follows we provide two approaches to convexify the optimization problem. 
	
	\subsection{Cost Modification Approach}
	First, we convexify the stability constraint by a change of variable and rewrite the cost function in \eqref{eq:quadratic_cost} in terms of this new variable. The following theorem formalizes the associated convex optimization problem obtained via this approach (which can be considered an approximation to the problem in \eqref{eq:nl_sdp_qp}). 
	
		\begin{thm}\textbf{\emph{(Stable Model Learning with Modified Cost)}}
		Consider system \eqref{eq:sys}, a given data-set of input and (estimated) uncertainty and state realizations. In addition, consider the uncertainty model of the form in \eqref{eq:model}. Consider the following convex program:
		\begin{mini!}|s|[2]                 	
	{P,S, R, W}                            
	{tr(W) \notag}   								
	{\label{eq:sdp_quadratic_cost}}             								
	{}                           								
	\addConstraint{A^\top P+P A+ S^\top +S}{\prec 0 \label{eq:stability_sdp_cost_modification}}   					  	
	\addConstraint{\left[\begin{array}{ccc}
			2P & \tilde{T}\tilde{D}^\top & I \\
			* & I  & {0} \\
			* & * & W
		\end{array}\right]}{\succeq 0 \label{eq:cost_related_constraint}}  						
	\addConstraint{P}{\succ 0 \notag}   					  	
\end{mini!}
	with given $A$ related to the known part of the system dynamics in \eqref{eq:sys}, $\tilde{T} := {\left[\begin{array}{ccc }
			S  & R & -P
		\end{array}\right]}$, and $\tilde{D}$ the Cholesky decomposition of the data matrix $D$ defined in \eqref{eq:data_matrix} (i.e., $D = \tilde{D}^\top \tilde{D}$). Denote the optimizers of \eqref{eq:sdp_quadratic_cost} as $P^\star$, $S^\star$, $R^\star$, and $W^\star$. Then, the following parameters of the model \eqref{eq:model}, $S_{\eta_l} = I, \Theta_l = \Theta_l^\star = P^{\star^{-1}} S^\star ,B_l = B_l^\star = P^{\star^{-1}} R^\star$ guarantee asymptotic stability of the model in \eqref{eq:model}. In addition, it holds that the cost $J$ of \eqref{eq:nl_sdp_qp} satisfies $J \leq tr(W)$; as such \eqref{eq:sdp_quadratic_cost} represents an approximate convexified problem of the problem in \eqref{eq:nl_sdp_qp}.
		\label{theorem:learning_modified_cost}
	\end{thm}
	\emph{\textbf{Proof}:} 
	The proof can be found in Appendix \ref{ap:learning_modified_cost}.
	\hfill $\blacksquare$
	
	\begin{rem} \emph{\textbf{(Surrogate Convex Optimization with Modified Cost)}}
	 We remark that the semi-definite program in \eqref{eq:sdp_quadratic_cost} is not equivalent to the non-convex optimization problem in \eqref{eq:nl_sdp_qp} (i.e., it is a convex approximation) due to setting $S_{\eta_l} = I$ and using a sufficient condition (a lower bound) in the derivation of the LMI in \eqref{eq:cost_related_constraint}. Although by letting $S_{\eta_l} = I$, we do not use the known structure of uncertainty, this makes the problem tractable. Note that here, we do not use knowledge of uncertainty structure. 
	\end{rem}
	
	Next, we follow a different approach to formulate an alternative surrogate (approximate) convex optimization problem for Problem \ref{prob:uncertainty_model_train}.

	\subsection{Constraint Modification Approach}
	
	Instead of changing the model-related variable ($\Theta_l$) in the stability constraint \eqref{eq:vdot_cond_Q} (or in its equivalent \eqref{eq:vdot_cond_P}), we formulate a sufficient condition (an upper bound) for the stability constraint \eqref{eq:vdot_cond_Q} which is linear in all the optimization parameters in order to convexify the optimization problem \eqref{eq:nl_sdp_qp}. The following theorem formalizes this approach. 
	
		\begin{thm}\textbf{\emph{(Stable Model Learning with Modified Constraint)}}
		Consider system \eqref{eq:sys}, a given data-set of input and (estimated) uncertainty and state realizations and the uncertainty model of the form in \eqref{eq:model}. Consider the following convex program:
		\begin{mini!}|s|[2]                 	
			{Q,\Theta_l, B_l, W}                            
			{tr(W) \notag}   								
			{\label{eq:sdp_quadratic_cost2}}             								
			{}                           								
			\addConstraint{\left[\begin{array}{cccc}
					A Q +Q A^\top   & S_\eta \Theta_l +\bar \gamma Q\\
				    \star & -2 \bar \gamma I\\
				\end{array}\right] }{\prec 0 \label{eq:stability_sdp_constraint_modification}}   					  	
			\addConstraint{tr(T D T^\top)}{\leq tr(W) \label{eq:cost_constraint_modification}}   					  	
			\addConstraint{Q}{\succ 0 \notag}   					  	
		\end{mini!}
			with given Hurwitz $A$, $S_\eta$ related to known parts of the system in \eqref{eq:sys}, positive scalar $\bar \gamma$, and $D$ and $T$ as defined in \eqref{eq:data_matrix}, and \eqref{eq:T}, respectively. Denote the optimizers of \eqref{eq:sdp_quadratic_cost2} as $Q^\star$, $\Theta_l^\star$, $B_l^\star$, and $W^\star$. Then, the following parameters of the model \eqref{eq:model}, $S_{\eta_l} = S_\eta, \Theta_l = \Theta_l^\star, B_l = B_l^\star$ guarantee asymptotic stability of the model in \eqref{eq:model}. In addition, it holds that the cost $J$ of \eqref{eq:nl_sdp_qp} satisfies $J \leq tr(W)$
		\label{theorem:learning_modified_constraint}
		\end{thm}
	\emph{\textbf{Proof}:} 
	The proof can be found in Appendix \ref{ap:learning_modified_constraint}.
	\hfill $\blacksquare$ \newline
	
Note that the scalar parameter $\bar \gamma$ in Theorem \ref{theorem:learning_modified_constraint} is tuned for the minimal feasible cost by a line search.

	\begin{rem} \emph{\textbf{(Surrogate Convex Optimization with Modified Constraint)}}
		Similar to Theorem \ref{theorem:learning_modified_cost}, the semi-definite program in \eqref{eq:sdp_quadratic_cost2} is a convex approximation of the non-convex optimization problem in \eqref{eq:nl_sdp_qp} since the stability constraint \eqref{eq:stability_sdp_constraint_modification} is a sufficient condition for asymptotic stability of the model in \eqref{eq:model}. Note that, a disadvantage of Theorem \ref{theorem:learning_modified_constraint} compared to Theorem \ref{theorem:learning_modified_cost} is that to ensure the feasibility of the semi-definite problem in Theorem \ref{theorem:learning_modified_constraint}, the known $A$ matrix of the system in \eqref{eq:sys} has to be Hurwitz. On the other hand, unlike Theorem \ref{theorem:learning_modified_cost}, Theorem \ref{theorem:learning_modified_constraint} uses the knowledge of uncertainty structure by setting $S_{\eta_l} = S_\eta$, which is potentially beneficial. 
	\end{rem}

In the above, we assumed that state and uncertainty realizations are available, which is, in practice typically not the case. In what follows, we present a solution for uncertainty and state estimation based on only input and output data.

\section{Uncertainty and State Estimation} \label{sec: uncertainty_state_estimation} 
	First, we formulate the uncertainty and state estimation problem before providing a solution for that problem. Consider system in \eqref{eq:sys} and the required assumptions as below to ensure that the problem is well-posed.

%
\begin{assum}[Regularity]\label{assumption1}
The following assumptions are required to ensure the regularity of the uncertainty and state estimation problem, which stand throughout this section: 
\begin{itemize}
	
	\item 
	\emph{\textbf{State and Input Boundedness:}}  The state variable $x_s(t)$ and the input $u(t)$ remain bounded in some compact region of interest. 
	
	\item \emph{\textbf{$\mathcal{C}^r$ Uncertainty Vector:}} The uncertainty vector $\eta(x_s(t), u(t))$ in \eqref{eq:sys} is $r$ times differentiable with respect to time, i.e., the time derivatives $\eta^{(1)}(x_s(t),u(t))$, $\eta^{(2)}(x_s(t),u(t))$, ... ,$\eta^{(r)}(x_s(t),u(t))$ exist and are continuous, and $\eta^{(r)}(x_s(t),u(t))$ is uniformly bounded.
	\label{assum:cr_assumption}

	\item \emph{\textbf{Disturbance Boundedness:}} The disturbance vector $\omega(t)$ in \eqref{eq:sys} is bounded uniformly in $t$.

	\item \emph{\textbf{$\mathcal{C}^1$ Measurement Noise:}} The measurement noise vector $\nu(t)$ in \eqref{eq:sys} is bounded uniformly in $t$ and differentiable, i.e., the total derivative with respect to time $\dot{\nu}(t)$ exists, is continuous, and bounded uniformly in $t$.
	
\end{itemize}

\end{assum}
	
%
%
%
	
	We assume input $u$ and measured output $y_s$ vector-valued signals in \eqref{eq:sys} are available. The following filter is designed for uncertainty and state estimation:
	\begin{equation} 		\label{eq:filter}
		\left\{\begin{aligned} 
			\dot{z} =&h(z,u,y_s;\theta), \\
			 \hat{\eta} =&\phi_1(z,y_s;\theta), \\
		    \hat{x}_s =&\phi_2(z,y_s;\theta),
		\end{aligned}\right.
	\end{equation}
	where $z \in {\mathbb{R}^{n_z}}$ is the internal state of the filter with ${n_z} \in \mathbb{N}$. Functions $h: \mathbb{R}^{n_z} \times \mathbb{R}^{l} \times \mathbb{R}^{m} \to \mathbb{R}^{n_z}$, $\phi_{1}: \mathbb{R}^{n_z} \times \mathbb{R}^{m} \to \mathbb{R}^{n_\eta}$, and $\phi_{2}: \mathbb{R}^{n_z} \times \mathbb{R}^{m} \to \mathbb{R}^{n}$ characterize the filter structure, $\theta$ denotes design parameters. 
	
	Define $\hat{x}_d := \text{col}[\hat{\eta}, \hat{x}_s]$ (representing the estimate of both the uncertainty and the state) and its estimation error as 
	\begin{equation} \label{eq:ed}
		e_d :=\hat{x}_d - x_d,
	\end{equation}
	where ${x}_d := \text{col}[{\eta}, {x}_s]$. The error dynamics of the filter is given later as a linear system and it is shown that $e_d = e_d(\omega,  \eta^{(r)},\nu, \dot{\nu})$. With this, we can state the uncertainty and state estimation problem at a high abstraction level. 

	\begin{prob}\emph{\textbf{(Uncertainty and State Estimation - Abstract Level)}} Consider the system \eqref{eq:sys} with known input and output signals, $u(t)$ and $y_s(t)$, and the uncertainty-state estimator filter \eqref{eq:filter}. For given $r$, design the filter parameters $\theta$ such that the following properties are guaranteed: \\
		\textbf{\emph{1) Stability:}} The trajectories of the filter \eqref{eq:filter} exist and are globally uniformly ultimately bounded for $t \geq 0$\emph{;} \\
		\emph{\textbf{2) Disturbance Attenuation:}} The $H_{\infty}$-norm of the transfer function from $\text{col}[\omega,  \eta^{(r)}]$ to $e_d$ in \eqref{eq:ed} is bounded by some known $\lambda > 0$\emph{;}\\
		\emph{\textbf{3) Noise Rejection:}} The $H_2$-norm of the transfer function from $\text{col}[\nu, \dot{\nu}]$ to $e_d$ in \eqref{eq:ed} is bounded by some known $\gamma > 0$.
		\label{prob:uncertainty_state_estimation_high_level}
	\end{prob}
	
	Now, to restate Problem \ref{prob:uncertainty_state_estimation_high_level} in a more formal way; first, we discuss the uncertainty-state estimator filter architecture, in what follows. 
	
	\subsection{Ultra Local Uncertainty Representation}
	First some preliminaries which are required to present the estimator filter architecture are discussed. Considering that the uncertainty $\eta(x_s(t), u(t))$ in \eqref{eq:sys} is an implicit function of time, for all $x_s(t)$ and $u(t)$, we can write an entry-wise $r$-th order Taylor time-polynomial approximation at time $t$ of $\eta$ as $\bar \eta = a_0 + a_1 t+ \dots + a_{r-1} t^{r-1}$ with coefficients $a_i \in {\mathbb{R}^{n_\eta}}, i= 0, \dots, r-1$. This model can be written in state-space form as
	\begin{equation} \label{eq:fault_model}
		\left\{\begin{aligned}
			\dot{\bar \zeta}_j &= \bar \zeta_{j+1},  \qquad 0 < j < r, \\
			\dot{\bar \zeta}_{r} &= 0,\\
			\bar \eta &= \bar \zeta_1,
		\end{aligned}\right.
	\end{equation}
	where $\bar \zeta_j \in {\mathbb{R}^{n_\eta}}$. Clearly, in the above model, we have $\bar \eta^{(r)} = {0}$, which might not be true for actual uncertainty signal $\eta$. Under Assumption \ref{assum:cr_assumption}, the actual internal state-space representation of $\eta$ is as follows:
	\begin{equation} \label{eq:fault_system}
		\left\{\begin{aligned}
			\dot{\zeta}_j &= \zeta_{j+1},  \qquad 0 < j < r, \\
			\dot{\zeta}_{r} &= \eta^{(r)},\\
			\eta &= \zeta_1,
		\end{aligned}\right.
	\end{equation}
	where $\zeta_j \in {\mathbb{R}^{n_\eta}}$. Clearly, the accuracy of the approximate model \eqref{eq:fault_model} increases as $\eta^{(r)}$ goes to zero (entry-wise). In the following, to design the uncertainty-state estimator we augment the system state, $x_s(t)$, with the states of the actual uncertainty internal state $\zeta_j(t), j \in\left\{1, \ldots, r\right\}$, and augment the system dynamics in \eqref{eq:sys} with \eqref{eq:fault_system}. We then design a linear filter (observer) for the augmented system to simultaneously estimate $x_s$ and $\zeta_j$ using model \eqref{eq:fault_model}. We remark that proper selection of the number of the uncertainty derivatives, $r$, added to the approximated model \eqref{eq:fault_model} (and \eqref{eq:fault_system}) is problem-dependent, see \cite{ghanipoor2022ultra} for discussion on selection of $r$.

	\subsection{Augmented Dynamics}
	Based on the uncertainty internal representation in \eqref{eq:fault_system} introduced above, define the augmented state $x_a:= \text{col}[x_s, \zeta_1,\zeta_2,\ldots,  \zeta_{r}]$, and rewrite the augmented dynamics using \eqref{eq:sys} and \eqref{eq:fault_system} as
	\begin{subequations} 		\label{eq:augmented}
		\begin{equation}\label{eq:augmented_system}
			\begin{aligned}
				\left\{\begin{aligned}
					\dot{x}_{a} &=A_{a} x_{a}+B_{u_a} u_a +B_{\omega_a} \omega_{a},\\
					y_s &= C_{a} x_{a}+D_\nu \nu,
				\end{aligned}\right.
			\end{aligned}
		\end{equation}
		\begin{equation} 		\label{eq:augmented_matrices}
			\begin{aligned}
				A_a &:= 
				\begin{bmatrix}
					A & S_\eta & {0} \\
					{0} & {0} & I_{d_n} \\
					{0} & {0} & {0} 
				\end{bmatrix}, \quad
				B_{u_a} := 
				\begin{bmatrix}
					B_u^\top & {0}
				\end{bmatrix}^\top, \quad u_a :=u, \\
			B_{\omega_a} &:= 
			\begin{bmatrix}
				B_\omega & {0} \\
				{0} & {0} \\
				{0} & I_{n_\eta}
			\end{bmatrix}, \quad 
			\omega_{a} :=	\left[\begin{array}{c}
				\omega \\
				\eta^{{(r)}} 
			\end{array}\right], \quad
				C_a := 
				\begin{bmatrix}
					C & {0}
				\end{bmatrix}
			\end{aligned}
		\end{equation}
	\end{subequations}
	with $d_n := (r-1)n_{\eta}$.
	\subsection{Uncertainty-State Estimator}
	In this section, considering the uncertainty-state estimator general structure in \eqref{eq:filter}, inspired from observer-based approaches, we consider $h(\cdot)$ and $\phi_i(\cdot), i=1,2$, as
	\begin{subequations} \label{eq:observer}
		\begin{equation}
			\begin{aligned} 		\label{eq:observer_dynamics}
				h(z,u,y;\theta) =& N z+G u+L y, \\
				\phi_i (z,y;\theta) =& \bar{C}_i (z-E y_S), \\
			\end{aligned}
		\end{equation}
		with $\hat{x}_a = z-E y_s$, filter state $z \in {\mathbb{R}^{n_z}}$, $n_z = n+r n_{\eta}$, 
		\begin{equation*} 
			\bar{C}_1 := {\left[\begin{array}{ccc}
					{0} & I_{n_\eta} & {0} 
				\end{array}\right]}, \quad
			\bar{C}_2 := {\left[\begin{array}{cc}
					I_n & {0}
				\end{array}\right]},
		\end{equation*}
		and matrices $(N,G,L)$ defined as
		\begin{equation}		\label{eq:observer_matrices}
			\begin{aligned}
				N &:=M A_a-K C_a,
				&M&:=I+E C_a, \\[1mm]
				G&:=M B_a,
				&L&:=K(I+C_a E)-M A_a E.
			\end{aligned}
		\end{equation}
	\end{subequations}
	Matrices $E$ and $K$ are filter gains to be designed which can be collected as $\theta = (E,K)$. Note that according to \eqref{eq:observer_dynamics}, the part of the augmented state, $x_a$, that we use to reconstruct uncertainty and state signals is $\bar{C}_a x_{a}$ with
	\begin{equation} 	\label{eq:c_bar_a}
		\bar{C}_a := {\left[\begin{array}{cc}
				\bar{C}_1^\top &
				\bar{C}_2^\top 
			\end{array}\right]^\top}.
	\end{equation}
	In the following section, we analyze the estimator error dynamics.  
	\subsection{Uncertainty-State Estimator Error Dynamics} 
	Consider the augmented state estimate $\hat{x}_a$ and let us define estimation error as 
	\begin{equation*}
		e:=\hat{x}_a-x_a=z-x_a-E y_s=z-M x_a-E D_\nu \nu.
	\end{equation*}
	Then, given the algebraic relations in \eqref{eq:observer_matrices}, the estimation error dynamics can be wrriten as
	\begin{equation*}
		\begin{aligned}
			\dot{e} &= Ne - M B_{\omega_a} \omega_{a} + {\left[\begin{array}{cc}
					K D_\nu & -E D_\nu
				\end{array}\right]} {\left[\begin{array}{cc}
					\nu \\ \dot{\nu}
				\end{array}\right]}.
		\end{aligned}
	\end{equation*}
	Define $\nu_a := \text{col}[\nu, \dot{\nu}]$, $e_d := \bar{C}_a e$ with $\bar{C}_a$ as in \eqref{eq:c_bar_a}, and
	$
	B_{\nu_a} := {\left[\begin{array}{cc}
			K D_\nu & -E D_\nu
		\end{array}\right]}.
	\label{eq:b_bar}
	$
	Then, the estimation error dynamics is given by
	\begin{equation}
		\left\{\begin{aligned}
			\dot{e} &=Ne - M B_{\omega_a} \omega_{a} + B_{\nu_a} \nu_a, \\
			e_d &= \bar{C}_a e.
		\end{aligned}\right.
		\label{eq:error_system}
	\end{equation}
	Define the transfer matrices
	\begin{equation}
		\begin{aligned}
			T_{e_d \omega_a}(s) &:= -\bar{C}_a(s I-N)^{-1} M B_{\omega_a}, \\
			T_{e_d\nu_a}(s) &:= \bar{C}_a(s I-N)^{-1} B_{\nu_a},
		\end{aligned}
		\label{eq:tfs}
	\end{equation}
	where $T_{e_d \omega_a}(s)$ and $T_{e_d\nu_a}(s)$, with $s \in \mathcal{C}$, denote the corresponding transfer matrices from $\omega_a$ and $\nu_a$, both to $e_d$, respectively. Now, we can restate Problem \ref{prob:uncertainty_state_estimation_high_level} in a more formal way. 
	
	\begin{prob}\emph{\textbf{(Uncertainty and State Estimation)}} Consider the system \eqref{eq:sys} with known input and output signals, $u(t)$ and $y_s(t)$. Furthermore, consider the internal uncertainty dynamics \eqref{eq:fault_system}, its Taylor approximation \eqref{eq:fault_model}, the augmented dynamics \eqref{eq:augmented}, the uncertainty-state estimator \eqref{eq:filter} with functions defined in \eqref{eq:observer}, and the transfer matrices in \eqref{eq:tfs}. Design the filter gain matrices $\theta = (E,K)$ such that the following properties are guaranteed: \\
		\textbf{\emph{1) Stability:}} There exist a class $\mathcal{K} \mathcal{L}$ function $\beta(\cdot)$ and a class $\mathcal{K}$ function $\mu(\cdot)$ such that for any initial estimation error $e(t_0)$ and any bounded input $\bar \omega_{a} :=\text{col}[\omega_a, \nu_a]$, the solution $e(t)$ of \eqref{eq:error_system} exists for all $t \geq t_{0}$ and satisfies
		\begin{equation*}\label{ISS_def}
			\|e(t)\| \leq \beta\left(\left\|e\left(t_{0}\right)\right\|, t-t_{0}\right) + \mu( \sup _{t_0 \leq \tau \leq t}\|\bar \omega_a (\tau)\| )  
	    \end{equation*}
		\emph{\textbf{2) Disturbance Attenuation:}} 
		\begin{equation} \label{eq:J1}
			J_1(\theta) = \|T_{e_d \omega_a}\|_{\infty} := \sup _{\mu \in \mathbb{R}^+} \sigma_{\max }(T_{e_d \omega_a}(i \mu))
		\end{equation}
		is bounded by some known $\lambda > 0$\emph{;} \\[1 mm]
		\emph{\textbf{3) Noise Rejection:}} 
		\begin{equation} \label{eq:J2}
			\begin{aligned}
				J_2(\theta) &= \|T_{e_d \nu_a}\|_{H_2} \\
				&:= \sqrt{\frac{1}{2 \pi} \operatorname{trace} \int_{-\infty}^{\infty} T_{e_d\nu_a}(i \mu) T_{e_d\nu_a}^{H}(i \mu) \mathrm{~d} \mu}
			\end{aligned}
		\end{equation}
		is bounded by some known $\gamma > 0$\emph{.}
		\label{prob:uncertainty_state_estimation}
	\end{prob}
	
	The essence of Problem \ref{prob:uncertainty_state_estimation} is to find an uncertainty-state estimator that, firstly,  ensures a bounded estimation error, $e(t)$, for any input (input to state boundedness); the $H_{\infty}$-norm of $T_{e_d \omega_a}(s)$, the transfer matrix from external disturbances and uncertainty model mismatch to the performance output $e_d = \bar{C}_a e$ is upper bounded by $\lambda$; the $H_2$-norm of $T_{e_d \nu_a}(s)$, the transfer matrix from measurement noise to the performance output $e_d$, is upper bounded by $\gamma$; and, for $\omega_{a} = {0}$ and $\nu_{a} = {0}$, $e(t)$ goes to zero asymptotically (internal stability).
	
	\subsection{Uncertainty-State Estimator Design}
	In the following proposition, we provide the solution of Problem \ref{prob:uncertainty_state_estimation} as a semi-definite problem, where we seek to minimize the $H_{\infty}$-norm of $T_{e_d \omega_a}(s)$ for an acceptable upper bound on the $H_2$-norm of $T_{e_d \nu_a}(s)$ (there exist a trade-off between these two norms, see \cite{1463317, khargonekar1996mixed}). Moreover, we add the Input-to-State Stability (ISS) constraint with respect to filter error dynamics input $\text{col}[\omega_a, \nu_a]$ to this program to enforce that stability of the resulting estimation filter. 
	
	\begin{prop}\emph{\textbf{(Estimator Design)}}
			Consider the system \eqref{eq:sys}, the augmented dynamics \eqref{eq:augmented}, the uncertainty-state estimator \eqref{eq:filter} with $h(\cdot)$ and $\phi(\cdot)$ as defined in \eqref{eq:observer}, the \linebreak corresponding estimation error dynamics \eqref{eq:error_system}, and the transfer functions \eqref{eq:tfs}. Consider the following convex program:
				\begin{mini!}|s|[2]                 	
				{\Pi, F, H, Z, \lambda, \gamma}                            
				{\lambda \notag}   								
				{\label{eq:uncertainty_state_estimator_design}}             								
				{}                           								
				\addConstraint{\bar S + \epsilon I}{\preceq 0  \notag}   					  	
				\addConstraint{\left[\begin{array}{ccc}
						\bar S & -(\Pi + F C_a) B_{\omega_a} & \bar{C}_a^{\top} \\
						* & -\lambda I  & {0} \\
						* & * & -\lambda I
					\end{array}\right]}{\prec 0  \notag}   					  	
				\addConstraint{\left[\begin{array}{ccc}
							\bar S & 	H D_\nu & -F D_\nu \\
						* & -\gamma I  & {0} \\
						* & * & -\gamma I
					\end{array}\right]}{\prec 0  \notag}
				\addConstraint{\left[\begin{array}{cc}
						\Pi & \bar{C}_a^{\top} \\
						* & Z
					\end{array}\right]}{\succ 0  \notag}
				\addConstraint{\Pi}{\succ 0  \notag}   					  	
				\addConstraint{\gamma - \operatorname{trace}(Z), \gamma, \lambda}{> 0  \notag}   					  	
				\addConstraint{\gamma }{\leq \gamma_{max}  \notag}
			\end{mini!}	
			with \\
			$\bar S := A_{a}^{\top} \Pi +A_{a}^{\top} C_{a}^{\top} F^{\top}-C_{a}^{\top}  H^{\top} + \Pi A_{a} +F C_{a} A_{a} -  H C_{a},$ \\
			given $\epsilon, \gamma_{max} > 0$, $\bar{C}_a$ in \eqref{eq:c_bar_a}, and the remaining matrices as defined in \eqref{eq:augmented_matrices}. Denote the optimizers as $\Pi^\star, F^\star, H^\star, Z^\star, \lambda^\star$, and $\gamma^\star$. Then, the optimal parameters in \eqref{eq:observer} \vspace{1 mm}\linebreak $\theta = \theta^\star = \{ E^\star = \Pi^{\star^{-1}} F^\star, K^\star = \Pi^{\star^{-1}} H^\star \}$ guarantee the following properties:
			\begin{enumerate}
				\item The estimation error dynamics in \eqref{eq:error_system} is ISS and the ISS-gain from input $\text{col}[\omega_a, \nu_a]$ to the estimation error is upper bounded by $ 2 \|\Pi^\star {\left[\begin{array}{ccc} (I+E^\star C_a) B_{\omega_a} & -K^\star D_\nu & E^\star D_\nu \end{array}\right]} \| \epsilon^{-1}$.
				\item $\|T_{e_d \omega_a}\|_{\infty}<\lambda^\star$ ($J_1(\cdot)$ in \eqref{eq:J1} is upper bounded by $\lambda^\star$).
				\item $	\|T_{e_d \nu_a}\|_{H_2}<\gamma^\star$ ($J_2(\cdot)$ in \eqref{eq:J2} is upper bounded by $\gamma^\star$).
		\end{enumerate}
		\label{theorem:optimal_estimator}
	\end{prop}
	\emph{\textbf{Proof}:} 
The proof follows the line of reasoning of the proof of Theorem 1 in \cite{ghanipoor2022linear}. 
\hfill $\blacksquare$


\section{Simulation Results}  \label{sec: sim results}
\begin{figure}[t!]
	\centering
	\smallskip
	\includegraphics[width=0.8\linewidth,keepaspectratio]{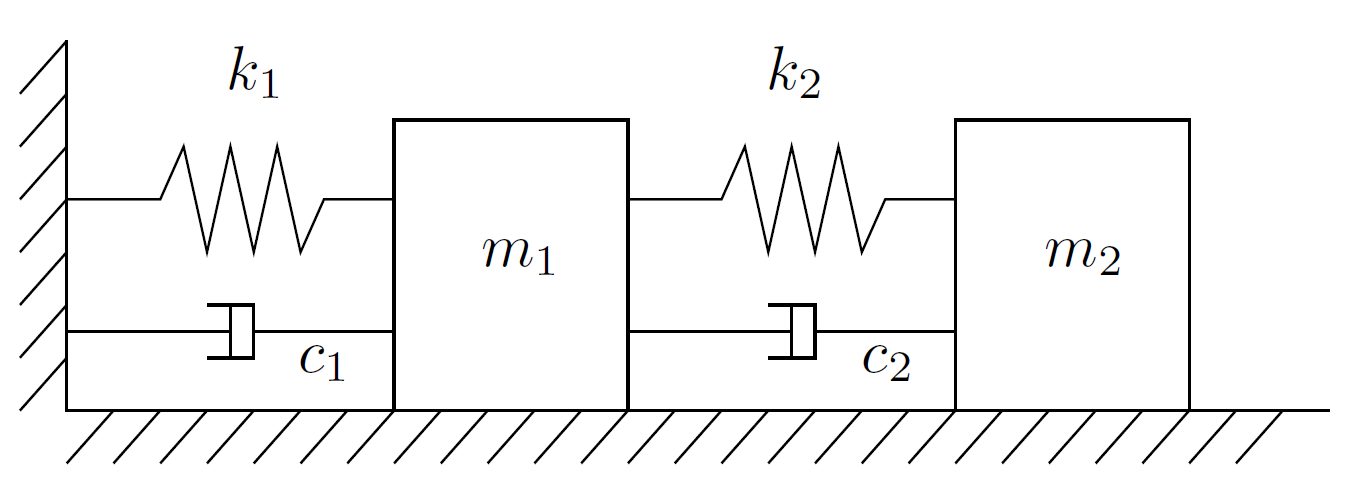}
	\caption{Two-mass-spring-damper schematic.}
	\label{fig:example}
\end{figure}

	In this section, we evaluate the proposed method using a two-mass-spring-damper system (see Fig. \ref{fig:example} for a schematic). By defining the state vector $x = [x_1, x_2, x_3, x_4]^\top := [q_1, \dot{q}_1, q_2, \dot{q}_2]^T$, where $q_i$ and $\dot{q}_i$ are the displacement and velocity of the $i-$th mass, respectively, the system dynamics can be described as follows:
\begin{equation}
	\left\{\begin{aligned}
		\dot{x}_s &= A x_s+ B_u u + S_\eta \eta \left(x_s\right), \\
		\eta \left(x_s\right) &= S_\eta \Theta_a x_s,   \\
		y_s &= C x_s + D_\nu \nu,
		\label{eq:example}
	\end{aligned}\right.
\end{equation}
where
\begin{equation*}
	\begin{aligned}
			A &=\left[\begin{array}{cccc}
				0 & 1 & 0 & 0 \\
				-\frac{k_1+k_2}{m_1} & -\frac{c_1+c_2}{m_1} & \frac{k_2}{m_1} & \frac{c_2}{m_1} \\
				0 & 0 & 0 & 1 \\
				\frac{k_1}{m_2} & \frac{c_1}{m_2} & -\frac{k_2}{m_2} & -\frac{c_2}{m_2}
			\end{array}\right], \quad 
			D= I, \\
			B_u &=\left[\begin{array}{c}
				0 \\
				 \frac{1}{m_1} \\
				  0 \\
				  0 \\
			\end{array}\right], \quad
		\Theta_a =\left[\begin{array}{cc}
		    -\frac{\delta k_1+\delta k_2}{m_1} & \frac{\delta k_1}{m_2} \\
		    0 & 0 \\
		    \frac{\delta k_2}{m_1} & -\frac{\delta k_2}{m_2} \\
		    0 & 0 \\
		\end{array}\right]^\top, \\
			S_\eta &=\left[\begin{array}{cccc}
				0 & 1 & 0 & 0 \\
				0 & 0 & 0 & 1 \\
			\end{array}\right]^\top, \quad
		C=\left[\begin{array}{cccc}
			1 & 0 & 0 & 0 \\
			0 & 0 & 1 & 0
		\end{array}\right], \\
		\end{aligned}
\end{equation*}
and constants $m_i$, $k_i$, and $c_i$ are the mass, stiffness, and viscous coefficient of the $i$-th mass, spring and damper, respectively. The uncertainty is due to the uncertainty on the springs stiffness which are captured by $\delta k_i$ for the $i$-th spring. Input $u$ is the force which applies to the first mass. The parameters values are: $ m_1 = 4$ $kg, m_2 = 3$ $kg, k_1 = 2$ $N/m, k_2 = 1.5$ $N/m, c_1 = 3.4$ $Ns/m, c_2 = 3.8$ $Ns/m, \delta k_1 = 0.25 k_1, \delta k_2 = -0.2 k_2$. For simulation, we set initial conditions as $x(0) = [0.01,0.01,0.01,0.01]^T$. 

For the above-mentioned system, $\Theta_a$ is unknown and we assume that we have a data-set of input, estimation of state and uncertainty (using uncertainty and state estimators discussed in Section \ref{sec: uncertainty_state_estimation}). Then, we want to solve Problem \ref{prob:uncertainty_model_train} (uncertainty model learning with stability guarantee) that finds an approximation of $\Theta_a$ as $\Theta_l$, using two proposed semi-definite programs in \eqref{eq:sdp_quadratic_cost} and \eqref{eq:sdp_quadratic_cost2}. Furthermore, we find uncertainty model parameter $\Theta_l$ without imposing stability constraint by only minimizing the quadratic cost function in \eqref{eq:quadratic_cost} for the training data-set. 

For this problem, given that $B_a$ is zero (i.e., the uncertainty is not a function of input signal), we have also selected $B_l$ as zero for all approaches.

\begin{figure}[t!]
	\centering
	\smallskip
	\includegraphics[width=1\linewidth,keepaspectratio]{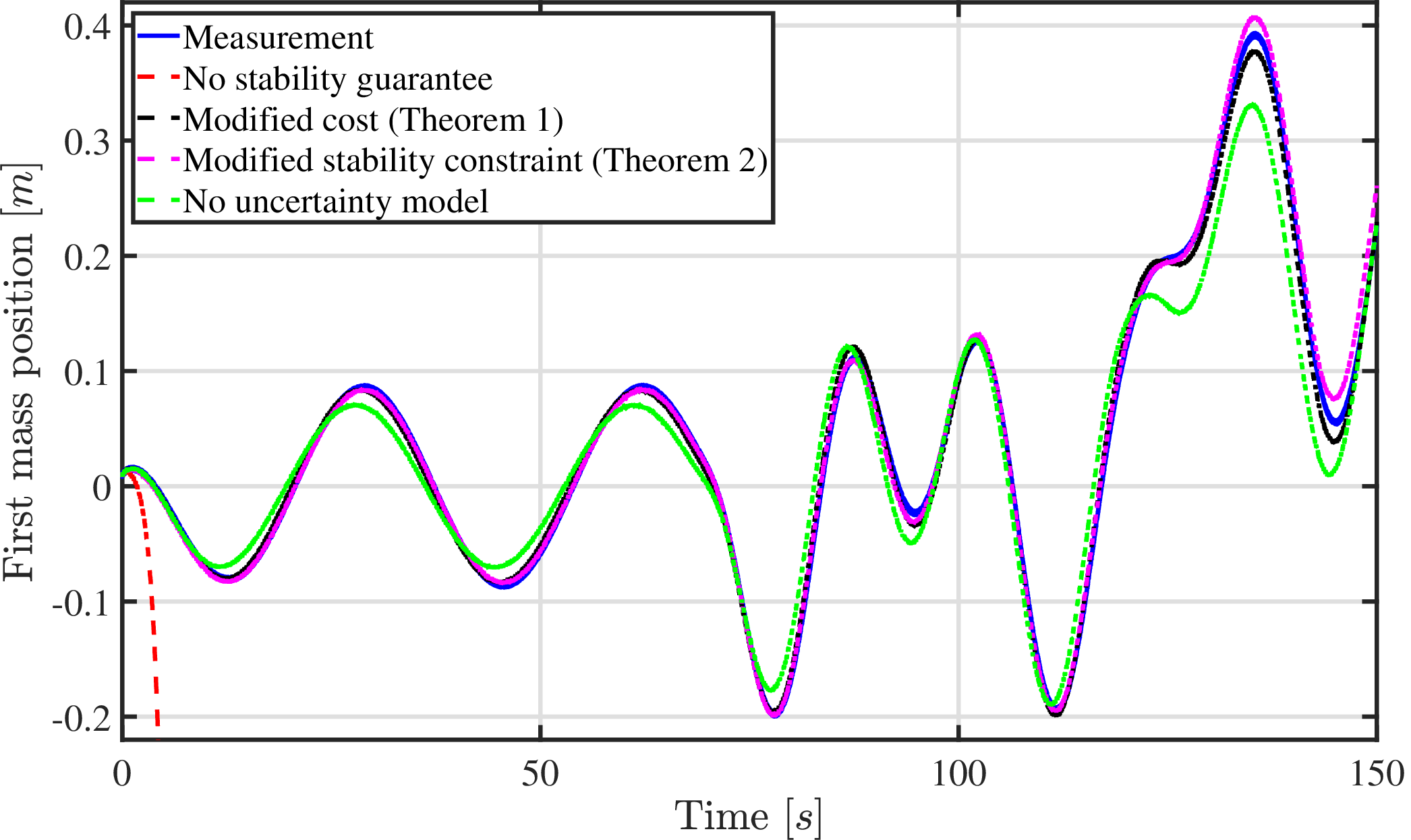}
	\caption{Comparison of first output of system and different models.}
	\label{fig:y1}
\end{figure}

\begin{figure}[t!]
	\centering
	\smallskip
	\includegraphics[width=1\linewidth,keepaspectratio]{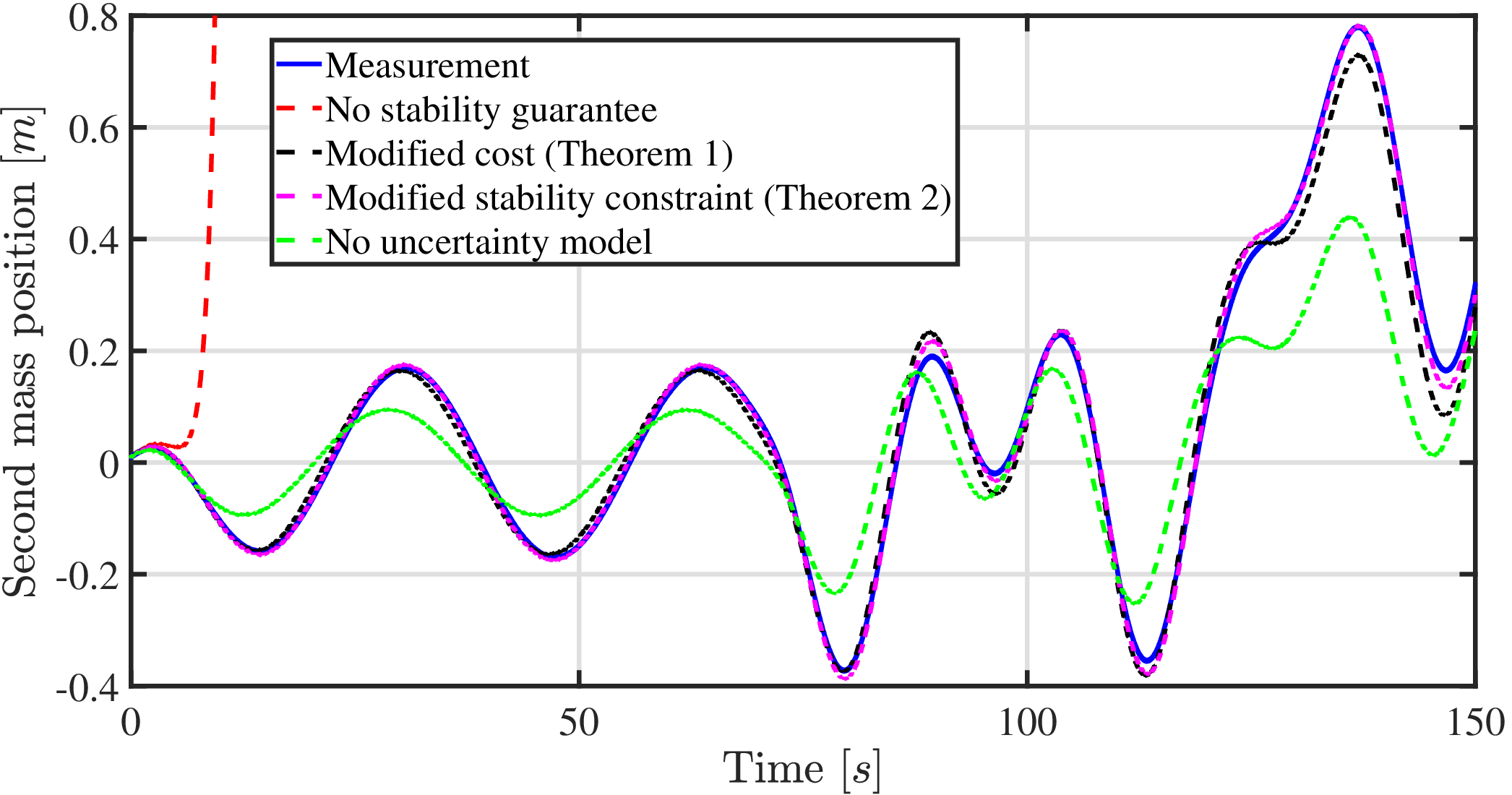}
	\caption{Comparison of second output of system and different models.}
	\label{fig:y2}
\end{figure}

For a test data-set, we have compared the output of system with extended models (which consist of the known model plus one of the uncertainty models) and also with the model without any uncertainty model. Figures \ref{fig:y1} and \ref{fig:y2} depict this comparison for the first and second mass positions, respectively. It can be seen from Figures \ref{fig:y1} and \ref{fig:y2}, that the result with uncertainty model which is trained without any stability constraint is unstable, see red dashed line. This shows that considering stability condition while learning a model for a stable system is indeed necessary.  Figures \ref{fig:y1} and \ref{fig:y2} also show that using the learning strategy proposed in this paper, model quality is significantly improved compared to the model without learned uncertainty model.

Furthermore, for better comparison, the Root Mean Square Errors (RMSEs) of error of each (stable) model (difference of model and system outputs) are given in Table \ref{tab:rmse}. As the results show, the constraint modification approach (Theorem \ref{theorem:learning_modified_constraint}) outperforms for this example. Note that we cannot generalize better performance of constraint modification approach in comparison with cost modification approach since we only show the results for one case study here.

\begin{table}[t!] 
	\caption{RMSEs of different models.}
	\centering
	\begin{tabular}{|c|c|c|}
		\hline
		RMSE [m]                                                           & \begin{tabular}[c]{@{}c@{}}First mass \\ position\end{tabular}  & \begin{tabular}[c]{@{}c@{}}Second mass\\ position\end{tabular}   \\ \hline
		 No uncertainty model &               $0.0296$                                                                      &   $0.1272$                                                                                                                      \\ \hline
		\multicolumn{1}{|c|}{\begin{tabular}[c]{@{}c@{}} Cost modification approach \\ \end{tabular}} &                          $0.0085$                                                            &   $0.0303$                                                                                                                                                \\ \hline
		\multicolumn{1}{|c|}{\begin{tabular}[c]{@{}c@{}}Constraint modification approach \\ \end{tabular}} &                          $0.0077$                                                            &   $0.0117$                                                                                                                                               \\ \hline
	\end{tabular} \label{tab:rmse}
\end{table}

\section{Conclusion}\label{sec: conclusion}

This paper proposes a framework for learning of modeling uncertainties in linear time invariant models. Furthermore, we guarantee stability of the extended model which includes a known physics-based model and the learned uncertainty model. We tackle this problem in two distinct steps: 1. we make the initial assumption that uncertainty and state estimation are already known, and under this assumption, we present two semi-definite programs (Theorems 1 and 2) to learn uncertainty models while guaranteeing asymptotic stability of the extended model; 2. we provide a filter, which can be designed using the semi-definite program in Proposition 1, to estimate uncertainty and state, given the known physics-based model and input-output data. Simulations for a two-mass-spring-damper system illustrate the proposed approach's performance and potential. Future work could include extension of the proposed method for a class of nonlinear systems. 

\appendix
	\section{Appendices}
\subsection{Proof of Theorem 1}  \label{ap:learning_modified_cost}  
We show that the non-convex optimization in \eqref{eq:nl_sdp_qp} can be convexified as \eqref{eq:sdp_quadratic_cost}. Consider the stability constraint \eqref{eq:vdot_cond_P}. First, to make the problem tractable, we set $S_{\eta_l} = I$. 

By applying the change of variable as $S := P \Theta_l$, the stability constraint in \eqref{eq:vdot_cond_P} is equivalent to \eqref{eq:stability_sdp_cost_modification}. Favorably, by the introduced change of variable, the stability constraint becomes convex (an LMI). However, since the cost $J$ in \eqref{eq:quadratic_cost} is a function of $\Theta_l = P^{-1} S$, the cost is not convex in $S$ and $P$, after the change of variable. Therefore, we have to convexify the cost to arrive at a convex optimization problem formulation. For scalar cost function $J$ in \eqref{eq:quadratic_cost}, we have the following
$$
J = \sum_{i = 1}^{N} d_i^\top T^\top T d_i = \sum_{i = 1}^{N} tr(d_i^\top T^\top T d_i),
$$
where $tr(\cdot)$ stands for trace operator. Due to cyclic property of the trace operator, the above cost can be written as  
$$
J = tr(T^\top T \sum_{i = 1}^{N} d_i d_i^\top). 
$$
Based on \eqref{eq:data_matrix} and the cyclic property of trace, we have 
$$
J = tr(T D T^\top).
$$
Now, we can write the epigraph form of the optimization problem in \eqref{eq:nl_sdp_qp} as follows:
\begin{mini}|s|[4]         	
	{P,S, B_l, W}                             
	{tr(W)}   								
	{\label{eq:optimization_in_thm1_proof}}             								
	{}                           								
	\addConstraint{A^\top P+P A+ S^\top +S}{\prec 0}   					  	
	\addConstraint{P}{\succ 0}   					  	
	\addConstraint{tr(T D T^\top)}{\leq tr(W).}   					  	
\end{mini}
Due to monotonicity of trace, the last constraint in \eqref{eq:optimization_in_thm1_proof} can be transformed to the following constraint:
\begin{equation*} 
	T D T^\top \preceq W,
\end{equation*}
By applying Schur complement to the above inequality, we have 
\begin{equation*} 
	\left[\begin{array}{cc}
		W & T\tilde{D}^\top \\
		\star & I
	\end{array}\right] \succeq 0.
\end{equation*}
Note that by construction, the data matrix $D$ is always symmetric and positive semi-definite. Therefore, its Cholesky decomposition (i.e., $D = \tilde{D}^\top \tilde{D}$) always exists. By applying the congruence transformation of $\text{diag}(P,I)$ to the above inequality, we obtain the following equivalent inequality
\begin{equation} \label{eq:GDG}
	\left[\begin{array}{cc}
		P W P &  \tilde{T} \tilde{D}^\top \\
		\star & I
	\end{array}\right] \succeq 0.
\end{equation}
Now, by substituting the lower bound of $2P-W^{-1}$ for $P W P$ and applying Schur complement, the LMI in \eqref{eq:cost_related_constraint} is obtained. In conclusion, instead of the non-convex optimization problem in \eqref{eq:nl_sdp_qp}, we provide an approximation of that in the form of the semi-definite program in \eqref{eq:sdp_quadratic_cost}. We remark that the approximation arises from initially setting $S_{\eta_l} = I$ at the beginning of the proof. Additionally, we use the lower bound of $2P-W^{-1}$ for $P W P$ in the derivation of the LMI in \eqref{eq:cost_related_constraint}.

\subsection{Proof of Theorem 2}  \label{ap:learning_modified_constraint}  
	Here, we follow the same line of reasoning as in the proof of Theorem 1 by showing that the non-convex optimization problem in \eqref{eq:nl_sdp_qp} can be convexified as in \eqref{eq:sdp_quadratic_cost2}. Consider the stability constraint \eqref{eq:vdot_cond_Q}. First, we set $S_{\eta_l} = S_\eta$ (i.e., here we use the knowledge of uncertainty structure). 
	
	Using Young's inequality $\big(X^\top Y+Y^\top X \preceq \frac{1}{2}(X+\bar Z Y)^\top \bar Z^{-1} (X+\bar Z Y)$, with a symmetric positive definite $\bar Z\big)$, we can find the following sufficient condition for the first inequality in \eqref{eq:nl_sdp_qp}:
\begin{equation*} 
		A Q +Q A^\top + \frac{1}{2}(\Theta_l^\top S_\eta^\top +\bar Z Q)^\top \bar Z^{-1} (\Theta_l^\top S_\eta^\top +\bar Z Q)  \prec 0.
\end{equation*}
Using the Schur complement, the above inequality is equivalent to 
\begin{equation*} 
	\left[\begin{array}{cc}
		A Q +Q A^\top  & S_\eta \Theta_l +Q \bar Z\\
		\star  & -2 \bar Z\\
	\end{array}\right] \prec 0.
\end{equation*}
Given nonlinear term $Q\bar Z$ in the above inequality we select $\bar Z = \bar \gamma I$, with a positive scalar $\bar \gamma$. Therefore, the above inequality can be written as \eqref{eq:stability_sdp_constraint_modification}. Furthermore, the quadratic cost can be treated as constraint \eqref{eq:cost_constraint_modification} by writing the epigraph form of the optimization problem (see \eqref{eq:optimization_in_thm1_proof} for the epigraph form). Thus, instead of the non-convex optimization problem in \eqref{eq:nl_sdp_qp}, we provide the surrogate semi-definite program in \eqref{eq:sdp_quadratic_cost2}.

	\section*{acknowledgment}
	This publication is part of the project Digital Twin project
	4.3 with project number P18-03 of the research programme
	Perspectief which is (mainly) financed by the Dutch Research
	Council (NWO).

	\bibliographystyle{IEEEtran}
	\bibliography{IEEEabrv,refs}

\end{document}